\documentclass[notitlepage,amssymb, nobibnotes,aps,nofootinbib,prd,showpacs,showkeys]{revtex4-1}

\usepackage{graphicx}
\usepackage{hyperref}
\usepackage{epsfig}
\usepackage{fancyhdr}
\usepackage{empheq}
\usepackage{mathbbol}
\usepackage{wasysym}
\usepackage{color}
\usepackage{bbm}
\usepackage{bm}
\usepackage{caption}
\usepackage{subcaption}

\newcommand{\Tr}{\text{Tr}}
\newcommand{\id}{1\!\!1}

\begin{document}

\title{On the color structure of Yang-Mills theory with static sources in a periodic box}
\author{L. Giusti$^a$, A.~L.~Guerrieri$^b$, S. Petrarca$^c$, A. Rubeo$^d$, M. Testa$^c$}
\affiliation{
\mbox{$^a$ Dipartimento di Fisica and INFN, Universit\`a di Milano-Bicocca, 
Piazza della Scienza 3, I-20126 Milano, Italy}\\
\mbox{$^b$ Dipartimento di Fisica and INFN, Universit\`a di Roma Tor Vergata, 
Via della Ricerca Scientifica 1, I-00133 Roma, Italy}\\
\mbox{$^c$ Dipartimento di Fisica and INFN, Universit\`a di Roma La Sapienza, 
P.le Aldo Moro 5, I-00185 Roma, Italy}\\
\mbox{$^d$ School of Mathematics, Trinity College, Dublin 2, Ireland}
\vspace{0.25cm}
} 

\begin{abstract}
We present an exploratory numerical study on the lattice of the color structure of the wave 
functionals of the SU(3) Yang-Mills theory in the presence of a $q\bar q$ static pair. In a spatial box with 
periodic boundary conditions we discuss the fact that all states contributing to the Feynman propagation 
kernel are global color singlets. We confirm this numerically by computing the correlations of 
gauge-fixed Polyakov lines with color-twisted boundary conditions in the time direction. 
The values of the lowest energies in the color singlet and octet external source sectors agree within 
statistical errors, confirming that both channels contribute to the lowest  (global singlet) state of the Feynman kernel. 
We then study the case of homogeneous boundary conditions in the time direction for which the gauge-fixing 
is not needed. In this case the lowest energies extracted in the singlet external source sector agree with 
those determined with periodic boundary conditions, while in the octet sector the correlator is 
compatible with being null within our statistical errors. Therefore consistently only the singlet external source 
contribution has a non-vanishing overlap with the null-field wave functional.
\end{abstract}

\maketitle

\section{Introduction\label{introduction}}
The knowledge of the energy levels containing infinitely heavy external sources, immersed in 
the quantum state of a pure Yang-Mills theory, is useful to understand the full QCD 
case~\cite{Brown:1979ya, McLerran:1981pb, Nadkarni:1986as, Bodwin:1994jh, Shuryak:2004tx, balipineda,Philipsen:2002az,Jahn:2004qr,Philps1}.
In Ref.~\cite{rostes1} the structure of the states in presence of external sources is analyzed on general 
grounds, within the temporal gauge, based on the formulation of Refs~\cite{rostes2, rostes3, rostes4, Rossi:2013iba}. 
In particular it is shown how to define projectors on states which are irreducible representations of 
the global color symmetry, and on states where the external sources are in a given color singlet or octet 
representation. The last ones are not, in general, eigenstates of the Hamiltonian. This fact led to  
some confusion in the literature, since a state with a definite global color behavior has non-trivial 
projections on both singlet and octet external source sectors.\\
\indent It is one of the aims of this paper to clarify this issue: we show that, when the theory is quantized 
in a finite space volume with periodic boundary conditions, Gauss's law enforces all states to be global 
color singlets. We also study in detail the properties of the ground state in presence of a static 
$q \bar q$ pair, and find that its wave functional projects both on the quark-singlet and on the 
quark-octet external source sectors. 
When homogeneous boundary conditions in the time direction are chosen, the lowest energies extracted in the 
singlet external source sector agree with those determined with periodic boundary conditions, while in the 
octet sector the correlator is consistently compatible with being null within statistical errors.\\
\indent The observables considered in this paper are correlators of Polyakov lines; it is well known that 
their signal to noise ratio is exponentially suppressed with the time-length of the lattice and with the spatial
distance of the sources. For the pure gauge theory
several techniques have been developed to overcome this problem for gluonic observables such as 
Polyakov loops, glueball interpolating operators, etc. 
\cite{Parisi:1983hm,Luscher:2001up,Luscher:2002qv,Meyer:2003hy,DellaMorte:2010yp}. 
In order to have a good statistical signal at large time distances, we opted  
for the multilevel algorithm proposed in Refs.~\cite{Parisi:1983hm,Luscher:2001up,Luscher:2002qv}. \\
\indent The paper is organized as follows. In section~\ref{theory} we collect the main aspects of the continuum theory:
we summarize the results of Ref.~\cite{rostes1}, we discuss the r\^ole of the Gauss's law, we introduce the 
"color-twisted" and 
homogeneous boundary conditions and we collect the formulas relevant for the analysis 
of the numerical data. Section~\ref{latttranscs} is devoted to the lattice transcriptions 
of the previous formulas. The results of our simulations, in part reported in 
Ref.~\cite{Guerrieri:2014pta}, are discussed in section~\ref{results}, while in 
section~\ref{conclusions} we draw our conclusions.
 
\section{Continuum Theory\label{theory}} 
\subsection{External sources\label{theoryES}}
In order to compute the Feynman propagation kernel in a sector in which external $q$ and 
${\bar q}$ sources are present, we have to perform, in the continuum, the following 
operations~\cite{rostes2}:
\begin{enumerate}
\item compute the propagation kernel with $A_0=0$
\begin{eqnarray}
{\tilde K} ({\bf A}_2, {\bf A}_1;T) =\int_{{\bf A} ({\bf x},0)= {\bf A}_1({\bf x})}^{{\bf A} ({\bf x},T)= {\bf A}_2({\bf x})} 
{\cal D} {\bf A} \exp [-S_{YM} ({\bf A}, A_0=0)] \, , \label{ktilde}
\end{eqnarray}
as a functional of the initial ${\bf A}_1$ and final ${\bf A}_2$ gauge three-dimensional configurations. 
Eq.~(\ref{ktilde}) describes a theory in which Gauss's law is not enforced.
\item enforce Gauss's law in presence of $q{\bar q}$ sources, through the integration over three-dimensional 
gauge transformations $\Omega$
\begin{equation}
K ({\bf A}_2, s_2, r_2; {\bf A}_1,s_1, r_1;T) = \int_{{\cal G}_0} {\cal D} \Omega \, \Omega_{s_2 s_1}({\bf x}_q) \, 
\Omega^*_{r_2 r_1}({\bf x}_{\bar q}) {\tilde K} ({\bf A}_2^\Omega, {\bf A}_1;T)\, ,\label{proj1}
\end{equation}
where 
\begin{equation}
A_\mu^\Omega(x) \equiv \Omega^\dagger(x)\,  A_\mu(x)\, \Omega(x) + i\, \Omega^\dagger(x)\, \partial_\mu \Omega(x)\; . 
\end{equation}
In Eq.~(\ref{proj1}) the $*$ denotes the complex conjugate, while $s_1$, $s_2$, $r_1$ and $r_2$ denote the initial and 
final color states of the $q$ and the ${\bar q}$ sources, respectively. The domain ${\cal G}_0$ of the 
invariant measure integration is restricted to the topologically trivial gauge transformations which tend to $1$ at 
three-dimensional infinity. In an infinite volume this is the condition needed in order not to eliminate from the 
Feynman propagation kernel the states with non-trivial global color content.
\end{enumerate}
\indent It is important to notice that when the theory is defined in a finite spatial volume with periodic 
boundary conditions, Gauss's law implies that the total color 
charge inside the volume carried by the gluons and the external sources is equal to the flux of the 
chromoelectric field through the boundary of the volume itself, and therefore vanishes identically 
as a consequence of periodicity. The color non-trivial states are therefore not present in the eigenspectrum 
of the finite volume theory with periodic boundary conditions, and a drastically different finite-size set 
up is needed to decide whether colored states can be excited or if their presence is suppressed as a consequence 
of the physics of color confinement.\\

The kernel $K$ in Eq.~(\ref{proj1}) is related to the energy eigenvalues $E_k$ and the corresponding energy 
eigenfunctionals $\psi_k ({\bf A},s,r)$ by 
\begin{equation}
K ({\bf A}_2, s_2, r_2; {\bf A}_1,s_1, r_1;T) = \sum_k e^{-E_k T} \psi_k ({\bf A}_2, s_2, r_2) 
\psi_k^* ({\bf A}_1, s_1, r_1) \, . \label{spectralcond}
\end{equation}
Under a gauge transformation $\Omega\in {\cal G}_0$, the states in Eq.~(\ref{spectralcond}) 
transform as
\begin{equation}
 \psi_k ({\bf A}^\Omega, s, r) = (\Omega^\dagger ({\bf x}_q))_{ss'} \, (\Omega ({\bf x}_{\bar q}))_{r'r} \, 
\psi_k ({\bf A}, s', r') \label{transprop} \, ,
\end{equation}
and the naive scalar product of two wave functionals 
is infinite. It must be replaced by the Faddeev-Popov gauge-fixed one 
\begin{equation}
(\psi,\phi) = \int {\cal D} \mu_F ({\bf A}) \sum_{rs} \psi^*({\bf A},r,s)\, \phi({\bf A},r,s) \, , \label{scalaprod}
\end{equation}
where
\begin{equation}
{\cal D} \mu_F ({\bf A}) = \Delta_F ({\bf A}) \delta [F ({\bf A})] \delta {\bf A}\; ,
\end{equation}
and the Faddeev-Popov determinant $\Delta_F ({\bf A})$ is defined by  \cite{FP}
\begin{equation}
\Delta_F ({\bf A}) \int_{{\cal G}_0} {\cal D} \Omega \, \delta[F({\bf A}^\Omega)] = 1 \, .
\end{equation}
As a consequence of Eq.~(\ref{transprop}), the value of the scalar product in Eq.~(\ref{scalaprod}) is independent 
from the choice of the gauge fixing functional $F({\bf A})$. In the following we assume 
$F({\bf A}) = {\bf \nabla} \cdot {\bf A}$.

\subsection{Global color transformations and classification of states\label{sec:IIB}} 
A global color transformation ${V}$ acts on the states as a unitary operator
\begin{equation}
[{\cal U}(V) \psi]({\bf A},s,r) = V_{ss'} V^*_{rr'} \psi({\bf A}^V,s',r') \label{colorrot}
\end{equation}
which commutes with the Hamiltonian~\cite{rostes1}. If we re-organize the source indices in a matrix,
the energy eigenstates in presence of $q{\bar q}$ sources can be written as 
\begin{equation}
\psi ({\bf A},s,r) = [\phi ({\bf A}) I + \lambda^a \phi^a ({\bf A})]_{sr} \equiv \psi_{sr} ({\bf A}) \, ,
\end{equation}
where the traceless generators satisfy the normalization condition $\Tr[\lambda^a \lambda^b]=\delta^{ab}/2$, 
and the color rotations in Eq.~(\ref{colorrot}) can be represented as
\begin{equation}
[{\cal U}(V) \psi] ({\bf A}) \equiv \psi^V ({\bf A}) = V \psi ({\bf A}^V) V^\dagger = 
\phi ( {\bf A}^V) I + V \lambda^a V^\dagger \phi^a ({\bf A}^V)\; . \label{spinorbit}
\end{equation}
Eq.~(\ref{spinorbit}) shows explicitly that the action of a color rotation results from a composition of a 
contribution coming from the action of $V$ on the source indices and an ``orbital'' one 
coming from the transformation ${\bf A} \rightarrow {\bf A}^V$. 
\begin{table}[t!]
\begin{center}
\begin{tabular}{|c|c|c|c|c|}
\hline
State & wave functional & orbital properties & $\phi$ at the null field conf. & global rep\\
\hline
singlet-singlet & $\psi^{({\bf 1,1})}({\bf A};i,j)=\delta_{ij}\phi({\bf A})$ & 
$\phi({\bf A}^V)=\phi({\bf A})$ & $\phi({\bf 0})\neq 0$ & singlet\\
singlet-octet & $\psi^{({\bf 1, 8})}({\bf A};i,j)=\lambda^a_{ij}\phi^a({\bf A})$ & 
$\phi^a({\bf A}^V)=\phi^a({\bf A})$ & $\phi^a({\bf 0})\neq 0$  & octet\\
$\boldsymbol \alpha$-singlet & $\psi_{m}^{({\bf \boldsymbol\alpha,1})}({\bf A};i,j)=\delta_{ij}\phi_m({\bf A})$ 
& $\phi_m({\bf A}^V)=R^{\boldsymbol \alpha}_{mm^\prime}(V)\phi_{m^\prime}({\bf A})$ & $\phi_m({\bf 0})= 0$ & 
$\boldsymbol \alpha$\\
$\boldsymbol \beta$-octet & $\psi_{m}^{({\bf \boldsymbol\beta, 8})}({\bf A};i,j)=\lambda^a_{ij}\phi^a_m({\bf A})$&
$\phi^a_m({\bf A}^V)=R^{\boldsymbol \beta}_{mm^\prime}(V)\phi^a_{m^\prime}({\bf A})$ & $\phi^a_m({\bf 0})= 0$ 
& $\boldsymbol \gamma \subset \boldsymbol \beta \otimes {\bf 8}$\\[0.05cm]
\hline
\end{tabular}
\caption{Classification of the color states with static $q\bar q$ sources. Every state is 
labeled by the orbital and the source representations respectively. The symbols $\boldsymbol \alpha$ 
and $\boldsymbol \beta$ indicate generic non-trivial representations.}
\label{tab:states}
\end{center}
\end{table}
In Tab.~\ref{tab:states} we give the explicit classification of the color states.
Several possible quantities can be computed, giving information about the structure of these states.
In the following we classify them on the basis of the time boundary conditions.

\subsection{Time-periodic boundary conditions \label{observables}}
The simplest quantity to measure is the ``total trace'' over color source indices, i.e.
the correlator of two Polyakov lines 
\begin{equation}
\mathcal{K}(R,T)\equiv\int {\cal D} \mu_F ({\bf A}) \sum_{s,r} K ({\bf A}, s, r; {\bf A},s, r;T) = \sum_k d_k e^{-E_k(R) T} \, . 
\label{eq:K}
\end{equation}
In Eq.~(\ref{eq:K}) $k$ runs over all the eigenstates of the Yang-Mills theory containing $q\bar q$ sources in all 
possible color configurations (see Table~\ref{tab:states}); $d_k$ is the multiplicity of the 
$k$-th level, and $E_k(R)$ is the corresponding energy which is a function of the relative spatial distance 
$R=|{\bf x}_q - {\bf x}_{\bar q}|$ between the two sources. The quantity $ \sum_{s,r} K ({\bf A}, s, r; {\bf A},s, r;T)$ 
is gauge invariant, and in the continuum the integration over ${\bf A}$ requires a gauge fixing. 
Following Ref.~\cite{rostes1}, we define the singlet and octet projectors on the color 
indices of the external sources 
\begin{align}
&\Pi^{\boldsymbol{1}}_{s_2,r_2,s_1,r_1}=\frac {1}{3} \delta_{s_2 r_2} \delta_{s_1 r_1},\label{eq:singproj}\\
&\Pi^{\boldsymbol{8}}_{s_2,r_2,s_1,r_1}=2 \sum_a \lambda^a_{r_2 s_2} \lambda^a_{s_1 r_1}=
\delta_{s_2 s_1} \delta_{r_2 r_1}-\frac {1}{3} \delta_{s_2 r_2} \delta_{s_1 r_1},\label{eq:octproj}
\end{align}
in terms of which we can define
\begin{align}
\mathcal{K}_{\boldsymbol{1}}(R,T)&=\int {\cal D} \mu_F ({\bf A}) \sum_{s_2,r_2,s_1,r_1} \Pi^{\boldsymbol{1}}_{s_2,r_2,s_1,r_1} 
K ({\bf A}, s_2, r_2; {\bf A},s_1, r_1;T) \,, \label{singoletto}\\ 
\mathcal{K}_{\boldsymbol{8}}(R,T)&=\int {\cal D} \mu_F ({\bf A}) \sum_{s_2,r_2,s_1,r_1} \Pi^{\boldsymbol{8}}_{s_2,r_2,s_1,r_1} 
K ({\bf A}, s_2, r_2; {\bf A},s_1, r_1;T) \, .\label{ottetto}
\end{align}
The integrands in Eqs.~(\ref{singoletto}) and (\ref{ottetto}) are not gauge invariant and 
require gauge fixing.

\subsection{Color-twisted boundary conditions in time \label{colortwisted}} 
In order to isolate the contributions from the various irreducible representations of 
the global color group, we introduce the color-twisted boundary conditions in the 
time direction requiring\footnote{In Ref.~\cite{DellaMorte:2010yp} analogous boundary conditions 
were considered for the construction of the projectors onto irreducible representations of the 
global symmetry groups of the Yang--Mills theory.}  
\begin{equation}
{\bf A}({\bf x},T) = V^\dag {\bf A}({\bf x},0) V\; , \label{perioda}
\end{equation}
where $V$ belongs to the global color group SU(3). An interesting quantity is~\cite{rostes1}
\begin{equation}
\mathcal{K}(V,R,T)\equiv\sum_{s_1,s_2,r_1,r_2} 
\int {\cal D} \mu_F ({\bf A}) V_{s_1 s_2} V_{r_1 r_2}^* K ({\bf A}^V, s_2, r_2; {\bf A},s_1, r_1;T) 
= \sum_{\boldsymbol\alpha, k}  \chi_{\boldsymbol\alpha}(V)  e^{-E^{\boldsymbol\alpha}_k (R) T}\, ,
\label{eq:totaltwist}
\end{equation}
where $ \chi_{\boldsymbol\alpha}(V)$ denotes the character of $V$ in the color representation $\boldsymbol\alpha$ 
to which the $k$-th level belongs. The contribution from the states transforming as an irreducible 
representation $\boldsymbol\alpha$ of the global color group are then given by
\begin{equation}
\mathcal{K}_{\boldsymbol\alpha}(R,T) = \mbox{dim}_{\boldsymbol\alpha}\; \frac{
\int D V\, \chi^{*}_{\boldsymbol\alpha}(V)\, \mathcal{K}(V,R,T)}{\int D V}\,\; ,
\end{equation}
where $\mbox{dim}_{\boldsymbol\alpha}$ is the dimension of ${\boldsymbol\alpha}$. Another quantity of interest is 
the response of the propagation kernel to an ``orbital'' color rotation, defined as
\begin{eqnarray}
\bar{\mathcal{K}}(V,R,T)\equiv\sum_{s,r} \int {\cal D} \mu_F ({\bf A})\, K ({\bf A}^V, s, r; {\bf A},s, r;T) 
\, . \label{tracechar1}
\end{eqnarray}

\subsection{Homogeneous boundary conditions in time \label{homo}} 
In Refs.~\cite{rostes1,rostes3} it is observed that in presence of 
the homogeneous boundary conditions
\begin{equation}
{\bf A}({\bf x},T)={\bf A}({\bf x},0)={\bf 0}\label{homboun}
\end{equation}
only the singlet-singlet and singlet-octet states survive in the kernel because they are the only ones
with a non-vanishing overlap with the null-field wave functional. Under these boundary conditions correlators 
of Polyakov lines are gauge invariant, and  the octet projections of the kernel on the color external state 
representations can be computed without gauge fixing. We have, in this case,
\begin{equation}
\mathcal{K}^{(0)}(R,T) = \mathcal{K}^{(0)}_{\boldsymbol{1}}(R,T) + \mathcal{K}^{(0)}_{\boldsymbol{8}}(R,T)\,,\label{eq:K0}\\
\end{equation}
where 
\begin{align}
&\mathcal{K}^{(0)}_{\boldsymbol{1}}(R,T) \equiv \frac{1}{3}\sum_{s,r} K ({\bf 0}, s, s; {\bf 0},r, r;T) = 
\sum_k  |\phi_{k}(\mathbf 0)|^2 e^{-E^{{\bf 1}}_{k} (R) T}\,,\label{eq:P1K0}\\
&\mathcal{K}^{(0)}_{\boldsymbol{8}}(R,T) \equiv \sum_{s,r} K ({\bf 0}, s, r; {\bf 0},s, r;T)-
\frac{1}{3} \sum_{s,r} K ({\bf 0}, s, s; {\bf 0},r, r;T) = 8 \sum_{a,k} |\phi^a_k(\mathbf 0)|^2 
e^{-E^{{\bf 8}}_{k} (R) T}\label{eq:P8K0}\, ,
\end{align}
and the superscript $(0)$ denotes homogeneous boundary conditions.

\section{Lattice transcription of the observables\label{latttranscs}} 
On a lattice with periodic boundary conditions we are interested 
in the ratio of the various kernels, as defined in the previous section,
normalized to the analogous one without external sources. In the case of 
Eq.~(\ref{eq:K}) the ratio $W(R,T)$ is  
\begin{equation}
W(R,T) = \left\langle \Tr(P({\bf x}))\,\Tr(P^\dag({\bf y})) \right\rangle_T\, ,\qquad
R=|{\bf x}_q - {\bf x}_{\bar q}|
\label{eq:latK}
\end{equation}
where the Polyakov line $P({\bf x})$  is defined by 
\begin{equation}
P({\bf x})=\prod_{x_0=0}^{T-1} U_0({\bf x},x_0) \, , \label{polloop}
\end{equation}
and $U_\mu(x)$ is the link matrix located at the position 
$x$ pointing toward the positive $\mu$ direction. Under a gauge transformation 
the trace over the color index $\Tr(P({\bf x}))$ is invariant, and, on the lattice, 
gauge fixing is not required to compute $W(R,T)$. The correlators corresponding to the 
quantities in Eqs.~\eqref{singoletto} and \eqref{ottetto} are 
\begin{align}
W_{\boldsymbol{1}}(R,T)&= \frac{1}{3}\left\langle\Tr(P({\bf x})P^\dag({\bf y}))
\right\rangle_T\label{eq:latsing}\\
W_{\boldsymbol{8}}(R,T)&=\left\langle \Tr(P({\bf x}))\Tr(P^\dag({\bf y})) \right\rangle_T - 
\frac{1}{3}\left\langle\Tr(P({\bf x})P^\dag({\bf y}))\right\rangle_T\label{eq:latoct}\; ,
\end{align}
where it is understood that Coulomb's gauge on the time slice at $x_0=0$ must be
fixed, see for instance Ref.~\cite{Giusti:2001xf}.\\

Twisted boundary conditions in the temporal direction are introduced by requiring that 
\begin{equation}
U_{\mu}(\mathbf x, T)=V^\dag U_{\mu}({\mathbf x}, 0)V\; . \label{boundv}
\end{equation}
The correlators associated to the quantities $\mathcal{K}(V,R,T)$ and 
$\bar{\mathcal{K}}(V,R,T)$ in Eqs.~\eqref{eq:totaltwist} and \eqref{tracechar1} 
are defined as 
\begin{equation}
W(V,R,T) = \left\langle\Tr(VP({\bf x}))\,\Tr(V^\dag P^\dag({\bf y}))\right\rangle_{T,V} \, ,\qquad
{\overline W}(V,R,T) = \left\langle\Tr(P({\bf x}))\,\Tr(P^\dag({\bf y}))\right\rangle_{T,V} \, , 
\qquad  \label{kernnostro}
\end{equation}
where the subscript $V$ indicates that the path integral integration is performed in presence of 
color-twisted boundary conditions, Eq.~(\ref{boundv}).\\ 

Finally, homogeneous boundary conditions are defined by 
\begin{equation}
U_{0}({\bf x}, T)=U_{0}({\bf x},0)=\id\; ,  
\end{equation}
and the definitions of the lattice correlators 
$W^{(0)}(R,T)$, $W^{(0)}_{\boldsymbol{1}}(R,T)$, $W^{(0)}_{\boldsymbol{8}}(R,T)$ follow straightforwardly from Eqs.~\eqref{eq:K0},~\eqref{eq:P1K0} 
and~\eqref{eq:P8K0}.

\section{Numerical results} \label{results}
The SU(3) Yang-Mills theory is discretized with the Wilson action. 
Numerical computations are performed by standard Monte Carlo techniques  
alternating $1$ heat-bath with $5$ over-relaxation steps. In the following 
the effective energy extracted from a correlator is defined as 
\begin{equation}
{\cal E}(R,T)=\frac{1}{\Delta} \ln\left[\frac{W(R,T - \Delta)}{W(R,T)}\right] \;,
\end{equation} 
where $\Delta=4$ unless $T=8$ for which $\Delta=2$.

\subsection{Color-twisted boundary conditions} \label{subsecA}
\begin{figure}
\centering
\begin{subfigure}[hbtp]{0.49\textwidth}
\includegraphics[width=0.85\textwidth]{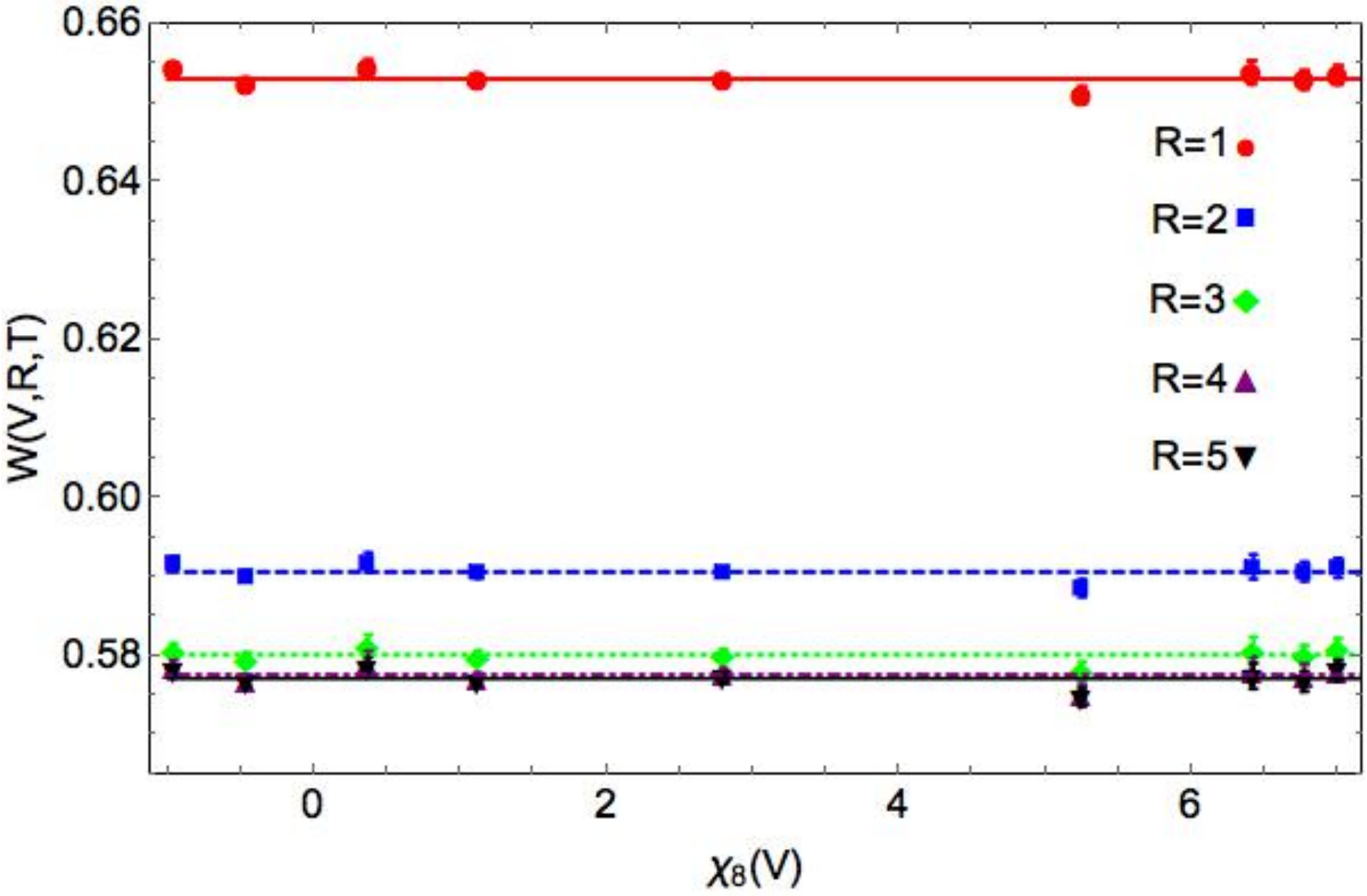}
\caption{}
\label{fig:TotalTwist}
\end{subfigure}
\begin{subfigure}[hbtp]{0.49\textwidth}
\includegraphics[width=0.85\textwidth]{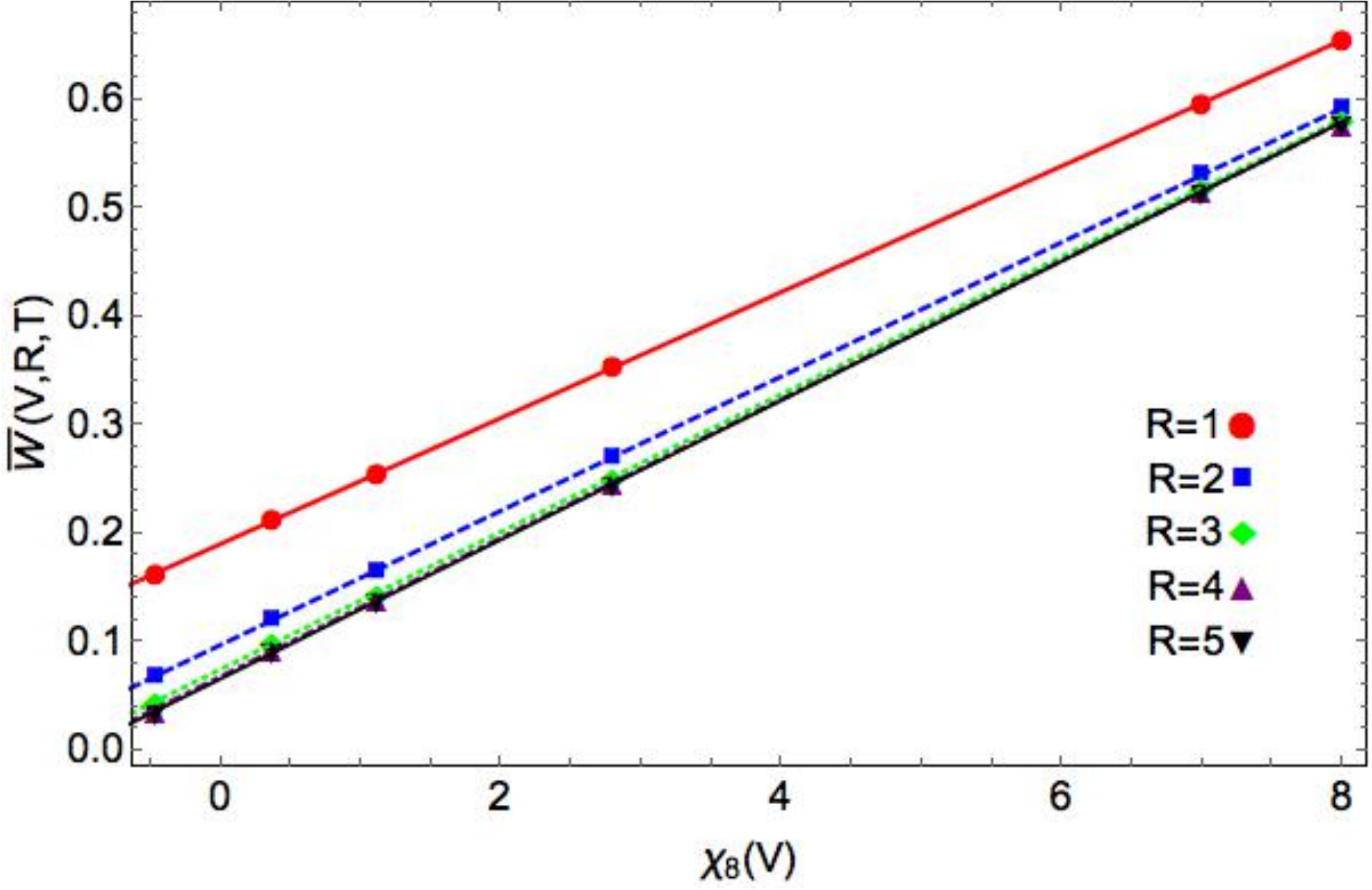}
\caption{}
\label{fig:PartTwistFit}
\end{subfigure}
\caption{$W(V,R,T)$ (a) and ${\overline W}(V,R,T)$ (b) as a function of the adjoint character $\chi_{\bf 8}(V)$.}
\end{figure}

We computed the dependence of $W(V,R,T)$ and $\bar W(V,R,T)$ defined in Eq.~(\ref{kernnostro}), 
at fixed $T$ and for different values of $R$, as a function of the octet character $\chi_{\boldsymbol{8}}(V)=|\Tr(V)|^2-1$. This run was 
carried out on a rather small volume, $10^3\times 4$ at $\beta=6.0$, in order to be able to keep under control the 
statistical noise without the necessity of using the multilevel technique. The number of generated configurations and 
the corresponding values of $\chi_{\boldsymbol{8}}(V)$ are listed in Tab.~\ref{tab:confs}.
\begin{table}[hbtp]
\begin{center}
\begin{tabular}{|c|c||c|c|}
\hline
$\chi_{\boldsymbol{8}}(V)$ & \# of confs. & $\chi_{\boldsymbol{8}}(V)$ & \# of confs.\\
\hline
$-0.96$ & $10000$ & $5.25$ & $4293$ \\
$-0.46$ & $16000$ & $6.41$ & $2453$ \\
$0.37$ & $4000$ & $6.77$ & $4000$\\
$1.12$ & $16000$ & $7.0$ & $4000$\\
$2.8$ & $16000$ & $8.0$ & $2000$\\
\hline
\end{tabular}
\end{center}
\caption{Number of configurations generated and the corresponding value of the adjoint character $\chi_{\boldsymbol{8}}(V)$ of the 
SU(3) matrix at the boundary. Notice that $-1<\chi_{\boldsymbol{8}}(V)\le 8$.}
\label{tab:confs}
\end{table}
The values of $W(V,R,T)$ are independent from the 
character $\chi_{\boldsymbol{8}}(V)$ at fixed values of $R$, see Fig.~\ref{fig:TotalTwist}.  
This is consistent with the expectation that in presence of periodic boundary conditions every state 
contributing to the propagation kernel is a global color singlet. The states in the octet external source 
sector, which contribute to the projected  
Feynman kernel in Eq.~\eqref{eq:latoct}, are therefore in an orbital-octet representation so to form 
an overall global singlet. This fact is numerically confirmed by the computation of 
${\overline W}(V,R,T)$, as shown in Fig.~\ref{fig:PartTwistFit} which exhibits the linear dependence on 
the adjoint character of $V$ typical of states in an orbital octet representation.

\subsection{Periodic boundary conditions\label{eq:pbcl}}
In order to compute the lowest energy eigenvalues of the states contributing to the correlators, the 
time extension $T$ of the lattice has been chosen large enough for the excited states to give a negligible 
contribution with respect to the statistical errors. For this purpose, $200$ independent configurations 
with $\beta=5.7$, volume $12^3\times T$, for $T=6, 8, 12, 16, 20$ have been generated. Each of these configurations
has been taken as the starting point for the multi-level averaging. The numerical values for the singlet and octet 
correlators $W_{\boldsymbol{1}}(R,T)$ and $W_{\boldsymbol{8}}(R,T)$ are reported in Tab~\ref{tab:SPK} and Tab.~\ref{tab:OPK} of the 
Appendix~\ref{app:appa}, respectively.
\begin{figure}
\centering
\begin{subfigure}[hbtp]{0.49\textwidth}
\includegraphics[width=0.85\textwidth]{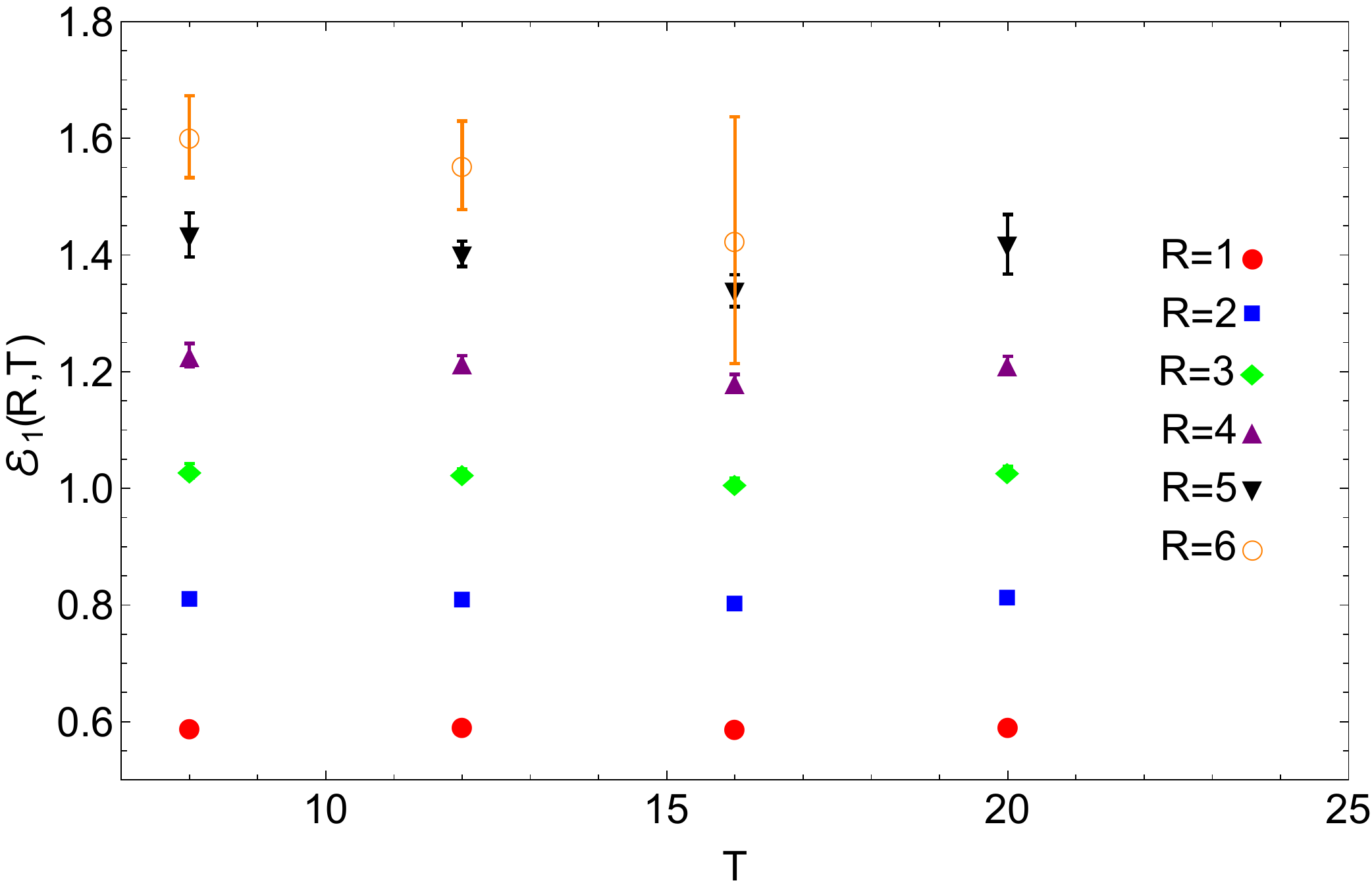}
\caption{}
\label{fig:P1K}
\end{subfigure}
\begin{subfigure}[hbtp]{0.49\textwidth}
\includegraphics[width=0.85\textwidth]{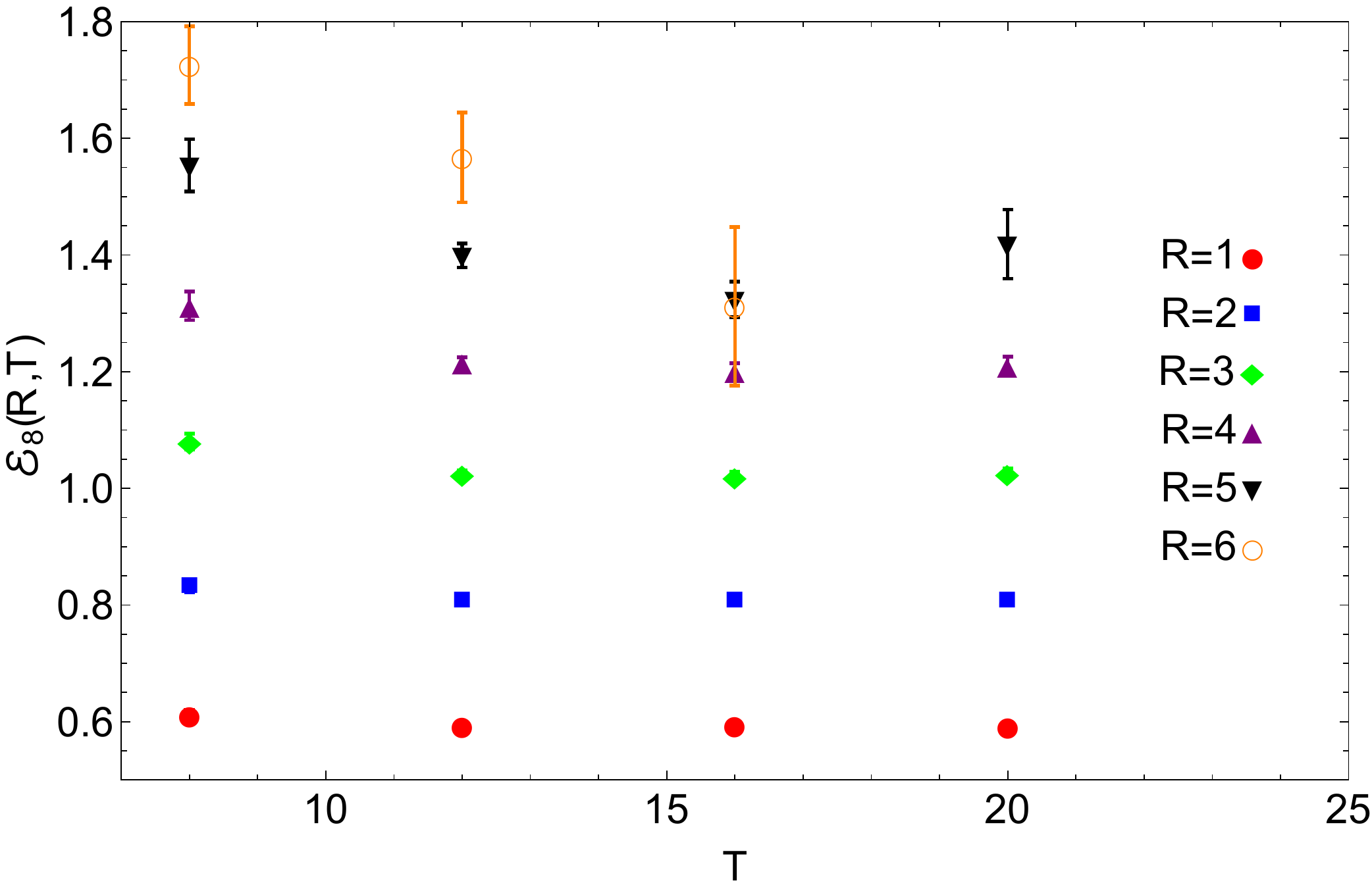}
\caption{}
\label{fig:P8K}
\end{subfigure}
\caption{Effective energies ${\cal E}_{\boldsymbol{1}}(R,T) $ (a) and ${\cal E}_{\boldsymbol{8}}(R,T)$ (b) as a  function of the temporal extension 
$T$ of the lattice for $R=1 \div 6$. The quantity $\Delta$ is equal to $4$ except for $T=8$ for which is $2$.}
\end{figure}
We computed the effective energies ${\cal E}_{\boldsymbol{1}}(R,T)$ and  ${\cal E}_{\boldsymbol{8}}(R,T)$
for the singlet and the octet correlator as a function of $T$ at a given separation $R$. 
The results are shown in Figs.~\ref{fig:P1K} and \ref{fig:P8K}, 
while their values at $T=20$ are reported in Table~\ref{tab:sing_oct_par}. 
\begin{table}[hbtp!]
\begin{center}
\begin{tabular}{|c|c|c|}
\hline
$R$ & ${\cal E}_{\boldsymbol{1}}(R,20)$ & ${\cal E}_{\boldsymbol{8}}(R,20)$ \\
\hline
 $1$ & $(5.93 \pm 0.02) \times 10^{-1}$ & $(5.91\pm0.05)\times 10^{-1}$ \\
 $2$ & $(8.10 \pm 0.05) \times 10^{-1}$ & $(8.07\pm 0.06)\times 10^{-1}$ \\
 $3$ & $1.029 \pm 0.009$ & $1.025\pm 0.010$ \\
 $4$ & $1.21 \pm 0.01$ & $1.21\pm 0.01$ \\
 $5$ & $1.42 \pm 0.05$ & $1.42 \pm 0.06$ \\
\hline
\end{tabular}
\end{center}
\caption{The values of ${\cal E}_{\boldsymbol{1}}(R,20)$ and  ${\cal E}_{\boldsymbol{8}}(R,20)$ as a function of the distance.}
\label{tab:sing_oct_par}
\end{table}
We assume the value of the effective energies computed at $T=20$ as our best 
numerical estimates.

We find that the singlet-singlet and octet-octet energies
coincide within statistical errors. This confirms that the ground state is a unique global color singlet with  
non-trivial components on the $q {\bar q}$ octet and singlet external source sectors. 

\subsection{Homogeneous boundary conditions}
As remarked above, the use of homogeneous boundary conditions in the time direction allows 
the computation of singlet and octet source index correlators without the need of gauge-fixing. 
\begin{figure}[hbtp!]
\begin{subfigure}[hbtp]{0.49\textwidth}
\begin{center}
\includegraphics[width=0.85\textwidth]{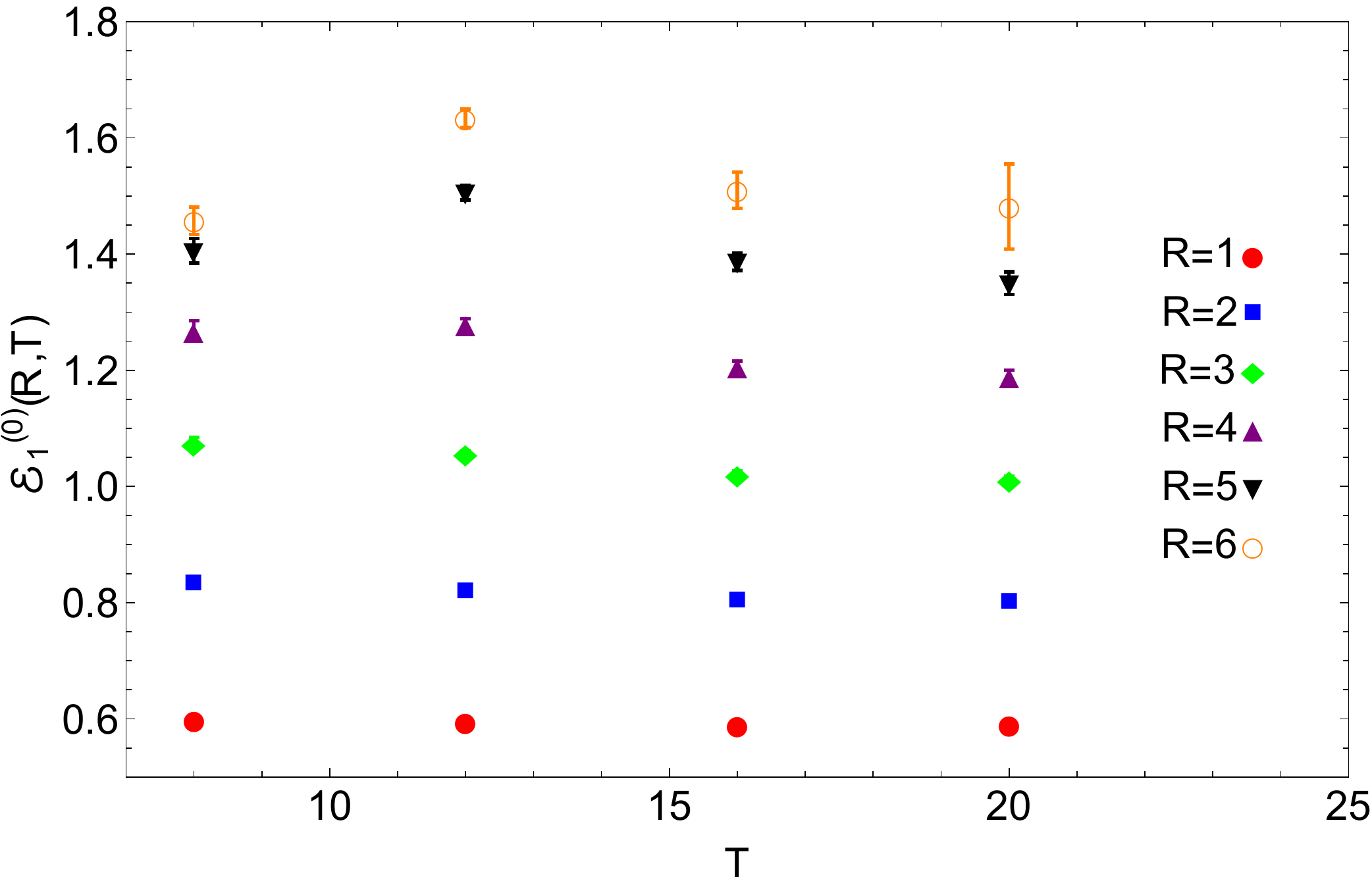}
\end{center}
\caption{}
\label{fig:P1K0}
\end{subfigure}
\begin{subfigure}[hbtp]{0.49\textwidth}
\begin{center}
\includegraphics[width=0.85\textwidth]{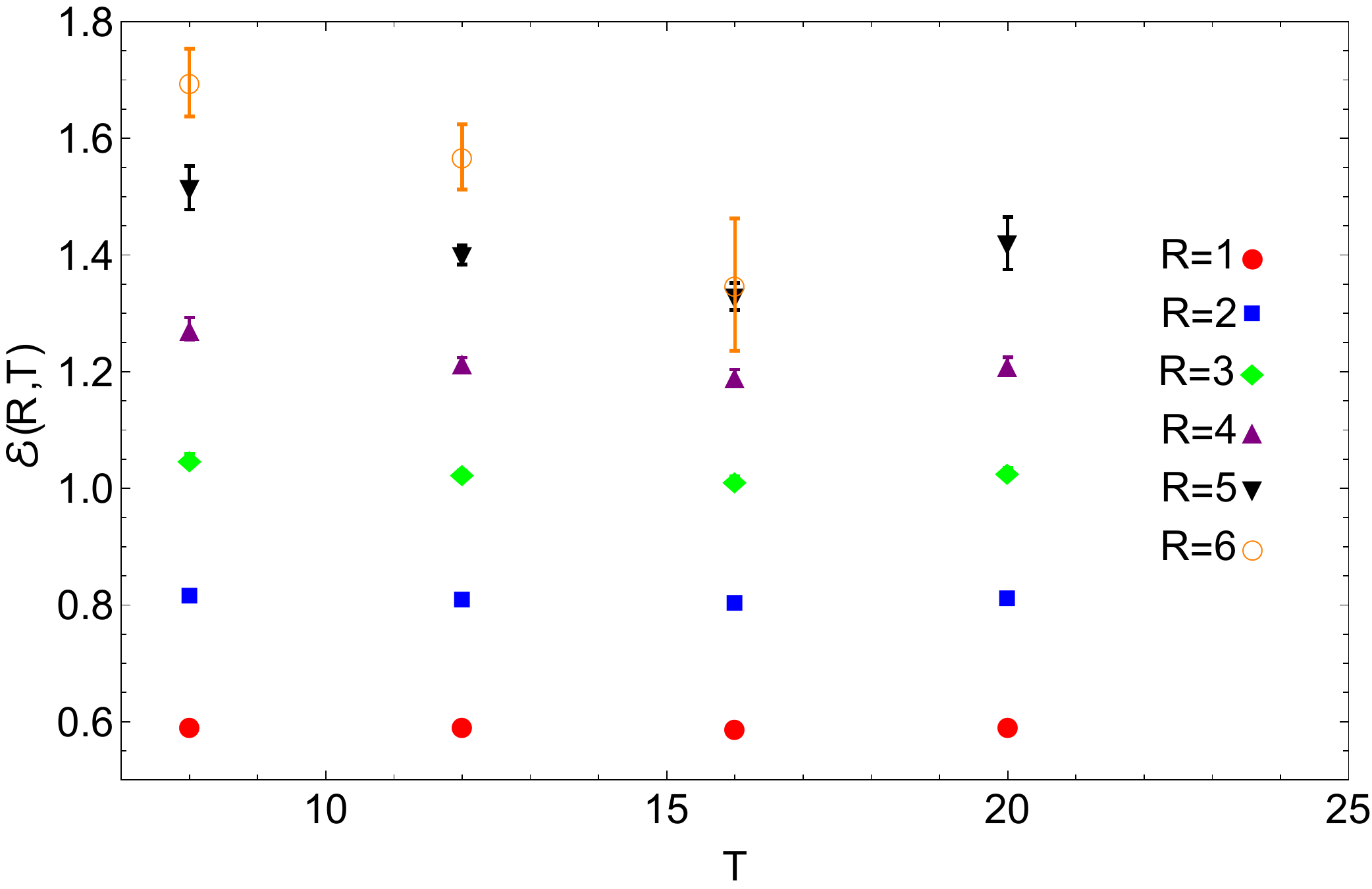}
\end{center}
\caption{}
\label{fig:K0}
\end{subfigure}
\caption{Effective energies ${\cal E}_{\boldsymbol{1}}^{(0)}(R, 20) $ (a) and ${\cal E}(R,20) $ (b) as a  function of the temporal 
extension $T$ of the lattice for $R=1 \div 6$. The quantity $\Delta$ is equal to $4$ except for $T=8$ for which is $2$.}
\end{figure}
As in section~\ref{eq:pbcl}, we generated $200$ configurations on which we applied the multilevel.
As compared with $W(R,T)$, $W^{(0)}_{\boldsymbol{1}}(R,T)$ turns out to have a larger pre-factor {as can be seen 
comparing the values of the kernel in Tab.~\ref{tab:SPKH} with those in Tab.~\ref{tab:SPK}}.
In Figs.~\ref{fig:P1K0} and \ref{fig:K0} we show the results for the effective energy $ {\cal E}_{\boldsymbol{1}}^{(0)}(R,T)$ of 
$W_{\boldsymbol{1}}^{(0)}(R,T)$ and for the effective energy ${\cal E}(R,T)$ of the gauge invariant $W(R,T)$ with periodic 
boundary conditions. The effective energies at $T=20$ are also reported in Table~\ref{tab:sing_FBC}. 

We find that the values extracted in the two cases are compatible within statistical errors, and that they are consistent  with those computed in the 
previous sub-section, see Table~\ref{tab:sing_oct_par}. The correlator $K^{(0)}_{\boldsymbol{8}}(R,T)$ is compatible 
with zero within statistical errors. This confirms the fact that only the singlet external source projection has a non-vanishing overlap with the null-field wave functional.
\begin{table}
\begin{center}
\begin{tabular}{|c|c|c|}
\hline
$R$ & ${\cal E}_{\boldsymbol{1}}^{(0)}(R,20)$& ${\cal E}(R,20)$\\
\hline
 $1$ & $(5.90 \pm 0.02) \times 10^{-1}$ & $(5.93 \pm 0.02)\times 10^{-1}$ \\
 $2$ & $(8.01 \pm 0.04) \times 10^{-1}$ & $(8.09 \pm 0.05)\times 10^{-1}$\\
 $3$ & $1.011 \pm 0.007$ & $1.028 \pm 0.008$ \\
 $4$ & $1.19 \pm 0.01$ & $1.21 \pm 0.01$ \\
 $5$ & $1.34 \pm 0.02$ & $1.42 \pm 0.04$ \\
\hline
\end{tabular}
\end{center}
\caption{The values of ${\cal E}^{(0)}_{\boldsymbol{1}}(R,20)$ and  ${\cal E}(R,20)$ as a function of the distance.}
\label{tab:sing_FBC}
\end{table}

\section{Conclusions}\label{conclusions}
In this paper we have discussed the color structure of the Yang-Mills states in the presence of  $q\bar q$ 
static quark sources with periodic boundary conditions in the spatial directions and periodic, color-twisted 
and homogeneous boundary conditions in the temporal one. A multilevel simulation algorithm has been necessary 
to obtain precise results on lattices corresponding to large times. From the outcome of our computations 
a general picture emerges for the structure of  states in sectors with external sources. The states contributing 
to the Feynman propagation kernel are global color singlets only. Our interpretation is that this fact does not necessarily have a dynamical meaning related to confinement, but is rather due to the periodic quantization conditions imposed in the spatial volume. We conclude that 
the states are of the form referred to as singlet-singlet or octet-octet. Of course, any mixing between these 
two cases is allowed, and only dynamical considerations can determine their relative weight in a quantitative way. 
Our numerical results are consistent with the observation that only the singlet external source contribution 
has a non-vanishing overlap with the null-field wave functional.

In view of the recent developments of the phenomenology of tetraquark (see for a review Ref. \cite{tetra}) and
pentaquark  \cite{penta} states we are undertaking the study of the non perturbative effective potentials acting in these channels, using the techniques described in the present paper.

\acknowledgements

We thank Giancarlo Rossi for many discussions and suggestions.

\appendix

\section{Numerical results for the correlators \label{app:appa}}
In this appendix we report the values of the correlators as a function of the distance between the 
external sources $R$ and the temporal extension $T$ computed with the multilevel algorithm. The errors 
are estimated using the jackknife method.
\begin{table}[hbtp]
\begin{center}
\begin{tabular}{|c|c|c|c|c|c|}
\hline
$R$ & $W_{\boldsymbol{1}}(R,6)$ & $W_{\boldsymbol{1}}(R,8)$ & $W_{\boldsymbol{1}}(R,12)$ &$W_{\boldsymbol{1}}(R,16)$ & $W_{\boldsymbol{1}}(R,20)$ \\
\hline
$1$  &     $(2.666 \pm  0.007)\times 10^{-2}$ & $(8.19 \pm 0.03)\times 10^{-3}$&$(7.72 \pm 0.03)\times 10^{-4}$ &$(7.25\pm 0.03)\times 10^{-5}$ &$(6.82\pm 0.03)\times10^{-6}$\\
$2$  &     $(6.28   \pm 0.04) \times 10^{-3}$  &$(1.254 \pm 0.009)\times 10^{-3}$&$(5.05 \pm 0.04)\times 10^{-5}$& $(2.02\pm0.02)\times10^{-6}$&$(8.04\pm0.09)\times10^{-8}$\\
$3$  &    $(1.42 \pm 0.02)\times 10^{-3}$ &$(1.82\pm 0.02)\times 10^{-4}$&$(3.11 \pm 0.04)\times 10^{-6}$&$(5.28\pm0.09)\times 10^{-8}$&$(8.86\pm0.18)\times10^{-10}$\\
$4$  &     $(3.78 \pm 0.06) \times 10^{-4}$ &$(3.29\pm 0.06)\times 10^{-5}$&$(2.68 \pm 0.06)\times 10^{-7}$&$(2.19 \pm 0.06)\times 10^{-9}$&$(1.79\pm0.06)\times 10^{-11}$\\
$5$  &     $(1.15 \pm 0.04)\times 10^{-4}$ &$(6.6\pm 0.2)\times 10^{-6}$&$(2.7 \pm 0.1)\times 10^{-8}$&$(1.03\pm0.07)\times 10^{-10}$&$(4.2\pm0.5)\times 10^{-13}$\\
$6$  &     $(6.3 \pm 0.4)\times 10^{-5}$ &$(2.5\pm 0.2)\times 10^{-6}$&$(6.0 \pm 0.9)\times 10^{-9}$&$(3.7\pm6.9)\times 10^{-12}$&$(5.1\pm4.9)\times10^{-14}$\\
\hline
\end{tabular}
\end{center}
\caption{Singlet projected correlator computed with the gauge-fixing and periodic boundary conditions.}
\label{tab:SPK}
\end{table}

\begin{table}[hbtp]
\begin{center}
\begin{tabular}{|c|c|c|c|c|c|}
\hline
$R$ & $W_{\boldsymbol{8}}(R,6)$ & $W_{\boldsymbol{8}}(R,8)$ & $W_{\boldsymbol{8}}(R,12)$ &$W_{\boldsymbol{8}}(R,16)$ & $W_{\boldsymbol{8}}(R,20)$ \\
\hline
$1$  &     $(2.42 \pm  0.02)\times 10^{-3}$    & $(7.11 \pm 0.06)\times 10^{-4}$ &$(6.63 \pm 0.06)\times 10^{-5}$  &$(6.22\pm 0.06)\times 10^{-6}$    &$(5.88\pm 0.06)\times10^{-7}$\\
$2$  &     $(1.92   \pm 0.02) \times 10^{-3}$  &$(3.64 \pm 0.03)\times 10^{-4}$  &$(1.44 \pm 0.01)\times 10^{-5}$  & $(5.74\pm0.07)\times10^{-7}$     &$(2.31\pm0.03)\times10^{-8}$\\
$3$  &    $(9.4 \pm 0.1)\times 10^{-4}$           &$(1.09\pm 0.01)\times 10^{-4}$   &$(1.82 \pm 0.02)\times 10^{-6}$  &$(3.04\pm0.05)\times 10^{-8}$     &$(5.18\pm0.10)\times10^{-10}$\\
$4$  &     $(4.8 \pm 0.1) \times 10^{-4}$         &$(3.56\pm 0.07)\times 10^{-5}$   &$(2.76 \pm 0.05)\times 10^{-7}$  &$(2.21 \pm 0.06)\times 10^{-9}$   &$(1.81\pm0.05)\times 10^{-11}$\\
$5$  &     $(2.7 \pm 0.2)\times 10^{-4}$          &$(1.27\pm 0.04)\times 10^{-6}$   &$(4.6 \pm 0.2)\times 10^{-8}$       &$(1.9\pm0.2)\times 10^{-10}$       &$(7.2\pm0.1)\times 10^{-13}$\\
$6$  &     $(2.1 \pm 0.1)\times 10^{-4}$          &$(7.3\pm 0.4)\times 10^{-6}$        &$(1.4 \pm 0.3)\times 10^{-8}$       &$(2.2\pm1.9)\times 10^{-11}$       & $-$\\
\hline
\end{tabular}
\end{center}
\caption{Octet projected correlator computed with the gauge-fixing and periodic boundary conditions.}
\label{tab:OPK}
\end{table}

\begin{table}[hbtp]
\begin{center}
\begin{tabular}{|c|c|c|c|c|c|}
\hline
$R$ & $W(R,6)$ & $W(R,8)$ & $W(R,12)$ &$W(R,16)$ & $W(R,20)$ \\
\hline
$1$  &     $(2.908 \pm  0.006)\times 10^{-2}$    & $(8.90 \pm 0.02)\times 10^{-3}$ &$(8.39 \pm 0.03)\times 10^{-4}$  &$(7.87\pm 0.03)\times 10^{-5}$    &$(7.41\pm 0.04)\times10^{-6}$\\
$2$  &     $(8.20  \pm 0.04) \times 10^{-3}$        &$(1.62 \pm 0.01)\times 10^{-3}$  &$(6.50 \pm 0.04)\times 10^{-5}$  & $(2.59\pm0.02)\times10^{-6}$     &$(1.04\pm0.01)\times10^{-7}$\\
$3$  &    $(2.36 \pm 0.02)\times 10^{-3}$           &$(2.92\pm 0.03)\times 10^{-4}$   &$(4.94 \pm 0.06)\times 10^{-6}$  &$(8.32\pm0.13)\times 10^{-8}$     &$(1.40\pm0.03)\times10^{-9}$\\
$4$  &     $(8.6 \pm 0.2) \times 10^{-4}$              &$(6.90\pm 0.11)\times 10^{-5}$   &$(5.43 \pm 0.09)\times 10^{-7}$  &$(4.4 \pm 0.1)\times 10^{-9}$       &$(3.59\pm0.10)\times 10^{-11}$\\
$5$  &     $(3.9 \pm 0.1)\times 10^{-4}$               &$(1.93\pm 0.06)\times 10^{-5}$   &$(7.3 \pm 0.3)\times 10^{-8}$       &$(3.0\pm0.2)\times 10^{-10}$       &$(1.1\pm0.1)\times 10^{-12}$\\
$6$  &     $(2.8 \pm 0.1)\times 10^{-4}$               &$(9.8\pm 0.5)\times 10^{-6}$        &$(2.1 \pm 0.3)\times 10^{-8}$       &$(2.6\pm2.0)\times 10^{-11}$       & $-$\\
\hline
\end{tabular}
\end{center}
\caption{Total trace correlator in presence of periodic boundary conditions.}
\label{tab:TPK}
\end{table}

\begin{table}[hbtp!]
\begin{center}
\begin{tabular}{|c|c|c|c|c|c|}
\hline
$R$ & $W_{\boldsymbol{1}}^{(0)}(R,6)$ & $W_{\boldsymbol{1}}^{(0)}(R,8)$ & $W_{\boldsymbol{1}}^{(0)}(R,12)$ &$W_{\boldsymbol{1}}^{(0)}(R,16)$ & $W_{\boldsymbol{1}}^{(0)}(R,20)$ \\
\hline
$1$  &     $(5.065 \pm  0.003)\times 10^{-2}$   & $(1.154 \pm 0.003)\times 10^{-2}$ &$(1.424 \pm 0.004)\times 10^{-3}$  &$(1.346\pm 0.005)\times 10^{-4}$  &$(1.265\pm 0.005)\times10^{-5}$\\
$2$  &     $(2.713  \pm 0.003) \times 10^{-2}$   &$(5.17 \pm 0.03)\times 10^{-3}$      &$(1.96 \pm 0.01)\times 10^{-4}$       & $(7.86\pm0.06)\times10^{-6}$     &$(3.15\pm0.03)\times10^{-7}$\\
$3$  &    $(1.684 \pm 0.003)\times 10^{-2}$      &$(1.97\pm 0.02)\times 10^{-3}$       &$(2.91 \pm 0.03)\times 10^{-5}$       &$(4.89\pm0.07)\times 10^{-7}$     &$(8.38\pm0.13)\times10^{-9}$\\
$4$  &     $(1.322 \pm 0.003) \times 10^{-2}$    &$(1.04\pm 0.02)\times 10^{-3}$       &$(6.33 \pm 0.11)\times 10^{-6}$       &$(5.01 \pm 0.10)\times 10^{-8}$       &$(4.19\pm0.10)\times 10^{-10}$\\
$5$  &     $(1.172 \pm 0.003)\times 10^{-2}$     &$(6.87\pm 0.16)\times 10^{-4}$       &$(1.70 \pm 0.04)\times 10^{-6}$       &$(6.4\pm0.2)\times 10^{-9}$       &$(2.8\pm0.1)\times 10^{-11}$\\
$6$  &     $(1.130 \pm 0.003)\times 10^{-2}$     &$(5.93\pm 0.16)\times 10^{-4}$        &$(8.5 \pm 0.3)\times 10^{-7}$           &$(1.7\pm0.1)\times 10^{-9}$       & $(5.1\pm0.8)\times 10^{-12}$\\
\hline
\end{tabular}
\end{center}
\caption{Singlet projected correlator computed in presence of homogeneous boundary conditions.}
\label{tab:SPKH}
\end{table}

\end{document}